\documentclass[review]{elsarticle}

\usepackage{hyperref}
\usepackage{graphicx}
\usepackage{amsmath}
\usepackage{amssymb}

\biboptions{authoryear}
\newcommand{\fref}[1]{Fig.~\ref{#1}}
\newcommand{\req}[1]{(\ref{#1})}

\begin{document}

\begin{frontmatter}

\title{Deformation Mechanics of Self-expanding Venous Stents: Modelling and Experiments}

\author[mymainaddress]{Masoud Hejazi}
\author[mymainaddress]{Farrokh Sassani}
\author[mysecondaryaddress]{J\"oel Gagnon}
\author[mysecondaryaddress]{York Hsiang}
\author[mymainaddress]{A. Srikantha Phani\corref{mycorrespondingauthor}}
\cortext[mycorrespondingauthor]{Corresponding author \newline phone: +1 (604) 822-6998) \newline fax: +1 (604) 822-2403 \newline email: srikanth@mech.ubc.ca}

\address[mymainaddress]{Department of Mechanical Engineering, 6250 Applied Science Lane, University of British Columbia, Vancouver, B.C, Canada V6T 1Z4}
\address[mysecondaryaddress]{Division of Vascular Surgery, 4219-2775 Laurel Street, Vancouver General Hospital, Vancouver, B.C, Canada V5Z 1M9}

\begin{abstract}
Deformation properties of venous stents  based on braided design, chevron design, Z design, and diamond design are compared using  \emph{in vitro} experiments coupled with analytical and finite element modelling. Their suitability for deployment in different clinical contexts is assessed based on their deformation characteristics.  Self-expanding stainless steel stents possess superior collapse resistance compared to Nitinol stents. Consequently, they may be more reliable to treat diseases like May-Thurner syndrome in which resistance against a concentrated (pinching) force applied on the stent is needed to prevent collapse. Braided design applies a larger radial pressure particularly for vessels of diameter smaller than $75\%$ of its nominal diameter, making it suitable  for a long lesion with high recoil. Z design has the least foreshortening, which aids in accurate deployment. Nitinol stents are  more compliant than their stainless steel counterparts, which indicates their suitability in veins. The semi-analytical method presented can aid in rapid assessment of topology governed deformation characteristics of stents and their design optimization.  


\end{abstract}


\end{frontmatter}


\section{Introduction}

Venous obstruction is a common pathological condition of the lower extremities, which reduces the vessel patency and hence the blood flow. Venous obstruction can be a result of a non-thrombotic syndrome (like May-Thurner and venous insufficiency) or an acute/chronic venous thrombosis~\citep{Murphy2017,beebe2005}. The prevalence and incidence rates of the venous obstruction vary depending on the underlying medical conditions. For instance, deep vein thrombosis (DVT) affects 300,000 people in North America each year with an incidence rate of $0.201\%$~\citep{Cohen2007,Silverstein1998,arshad2017}. Also, the prevalence rate of May-Thurner and chronic venous insufficiency are $<40\%$ and $<73\%$, respectively~\citep{cavalcante2015,beebe2005,radaideh2019}. Some recent studies support endovascular treatment (stenting) over medical/compressive therapy~\citep{rossi2018}. 

Vascular stents are small mesh-like porous tubular scaffolds, which are deployed inside diseased blood vessels to restore patency. These devices are available in various sizes, structures, and materials to provide desired mechanical properties in each particular design. Self-expanding and balloon-expandable stents are two major categories; each has a specific deployment procedure and expansion mechanism resulting in a different clinical performance. Since the balloon-expandable stents have a small range of elastic expansion, their self-expanding counterparts are preferred for deployment in  veins because of their higher compliance and the ability to retain patency during the physiological dilation of veins~\citep{Wittens2015}. Venous stenting evolved from the endovascular treatment of occlusive arteries. While the etiology of arterial and venous disease is different, arterial stents have been commonly used in off-label applications in venous endovascular treatment. However, recent studies on venous stenting suggested the need for designing venous stents accounting for the specific venous pathology~\citep{schwein2018,bento2019}. The design and manufacturing of venous stents have been overlooked and undervalued~\citep{gordon2008} given the prevalence of venous disease and the  off-label use of arterial stents. The main goal of this study is to examine stent deformation characteristics that contribute to their clinical performance and hence aid in the selection of a stent in a given clinical application. 

Although atherosclerosis is the main etiology for the arteries, a chronic venous disease occurs due to venous thrombosis and external compression by the adjacent artery~\citep{bento2019}. Compared to arteries, veins are up to three times more compliant (distensible). Pathological arteries usually retain a well-defined vessel with almost no change in the vessel elasticity~\citep{schwein2018}. However, a pathological vein can undergo a fibrous retraction reducing its compliance. During the early stages of thrombosis, the venous thrombus is compliant and differentiable from the venous wall. During the chronic phase, $23-60\%$ of acute DVT cases, the fibrotic thrombus is attached to the wall causing vein thickening and post-thrombotic recoil \citep{razavi2015,deatrick2011}. In summary, in designing a venous stent, one should consider the localized pinching forces (May-Thurner), high recoil (fibrotic veins), foreshortening (reduction of the length during expansion), and the large distensibility (high compliance) of the healthy wall proximal and distal to the pathological region.  We first  review pertinent literature on the deformation properties of stents.


The  effect of radial pressure on the hemodynamics of a stented vessel has been investigated through experimental and computational studies~\citep{Freeman2010,Bedoya2006,Lally2005,Zahedmanesh2009}. It is shown that a stent with excessive radial pressure constrains the cyclic dilation of the vessel leading to post-deployment complications such as restenosis~\citep{Morrow2005,Vernhet2001}. Nevertheless, an optimal radial pressure is required to avoid post-deployment recoil and migration~\citep{Li2006}. The radial pressure is a function of stent structural parameters and the material properties, which has been extensively studied for braided stents~\citep{zaccaria2020,Kim2008}, Z stents~\citep{SnoWhill2001}, and closed-cell venous stents~\citep{dabir2018}. In venous stenting, oversizing the stent is a common technique to avoid recoil and migration. Here we provide a semi-analytical method, in Section 3, to determine the radial pressure variation due to stent oversizing.


Collapse  is a common failure mode for stents that are placed in veins~\citep{Murphy2017}. Three different buckling modes in the collapse of balloon-expandable stents are identified in~\citep{Dumoulin2000} identified and collapse in self-expanding stents has been reported in~\citep{Kim2008,dabir2018,schwein2018}. They observed mechanical instability by applying a pinching force on the stent structure. It has been  reported that the local collapse stiffness of a stent depends critically on its strut geometry and the elastic modulus of the stent material~\citep{Duerig2000}. We evaluate the performance of a venous stent against collapse experimentally and explain how we can qualitatively compare a stent’s behavior under collapse  through a unit-cell study. 


Compliance of a stent also plays a vital role in its clinical performance particularly for the venous system with high distensibility. The stent is deployed to maintain adequate contact with the healthy wall proximally and distally in order to increase the anchoring area and avoid migration.  The compliance mismatch between the vessel and the stent magnifies the post-deployment complications~\citep{Berry2002,Morris2016,post2019}. The compliance of a stent/vessel is defined as the variation of pressure over the variation of diameter. Here, the stent compliance is determined through experiment and analysis.  


Despite the critical role of the foreshortening in precise stent placement, there is little study on the foreshortening of venous stents. For balloon-expandable arterial stents, however, the significance of longitudinal strain and the foreshortening mechanism has been elaborated in the literature~\citep{Douglas2014,Tan2011}. However, the method cannot be used here due to differences in the expansion mechanism of self-expanding stents. Here, we measure the foreshortening during stent expansion and analytically study this parameter in Section 3.  

The above studies confirm the importance of  mechanical properties and deformation characteristics on the clinical performance of venous stents and a need to develop rapid assessment tools to inform the choice of stent topologies. Deformation characteristics of venous stents  based on braided design, chevron design, Z design, and diamond design are compared using  \emph{in vitro} experiments coupled with analytical and finite element modelling. Their suitability for deployment in different clinical contexts is assessed based on their deformation characteristics.  We start with an  \emph{in vitro} experiment to evaluate these parameters in Section 2. Afterwards, we employ the unit-cell study in Section 3 to determine radial pressure, compliance, foreshortening and collapse of the candidate stents in this study. We assess the validity of the unit-cell study by comparing the results with  observations from the  \emph{in vitro} experiment and discuss clinical relevance in Sections 4 and 5.  Concluding remarks and future work are given in Section 6.

\section{\emph{In vitro} experiments}

\begin{figure}[!h]
	\centering
	\includegraphics[width=\textwidth]{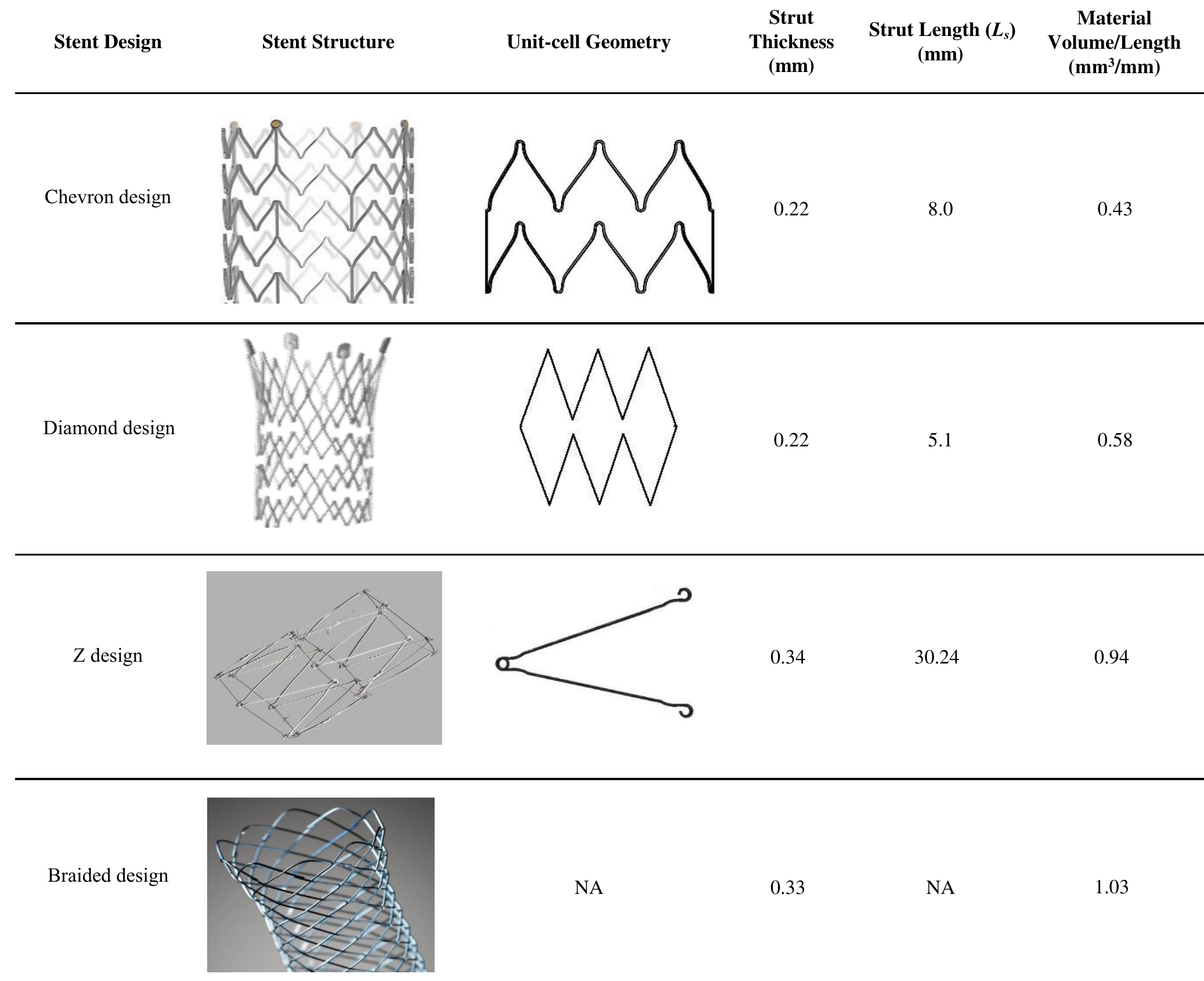}
	\caption{Stent designs used in this study and their structural design parameters.}
	\label{Fig_1}
\end{figure}

Self-expanding stents are available in different structural design and materials. Here we chose two stainless steel stents (Z design and Braided design) and two Nitinol stents (Diamond design, Chevron design) shown in~\fref{Fig_1}. Note that these stents are commercial designs currently being used in practice for venous stenting and here we use the design names instead of commercial names provided by companies. Among these, Braided design is the only one with an open-cell structure and the rest are closed-cell designs. Diamond design and Chevron design are the recent designs dedicated to venous stenting, Braided design is a common off-labeled design (used for both arteries and veins), and Z design is a trachea stent commonly used for venous stenting as a local reinforcement~\citep{Murphy2017}. To study the effect of the material and structural design, we start with a series of  \emph{in vitro} tests to determine the radial pressure, collapse resistance, and foreshortening. Compliance, defined as the diameter variation ratio to  radial pressure can be calculated as well.  

\begin{figure}[!h]
	\centering
	\includegraphics[width=\textwidth]{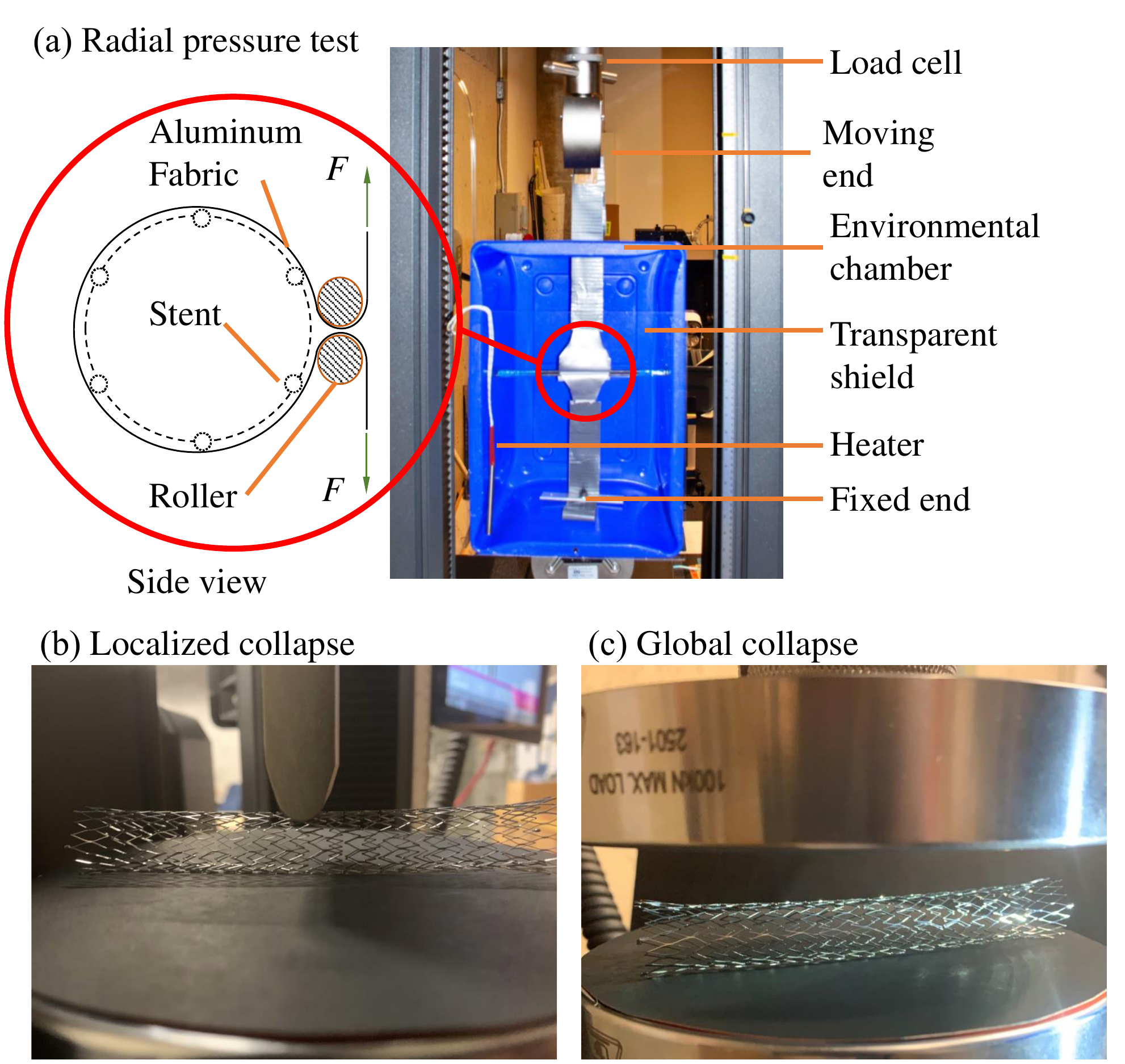}
	\caption{\emph{In vitro} tests. (a) Stent is wrapped in an aluminum fabric sheet that is attached to the material testing machine (Instron 5965) grips at both ends. The rollers reduce the friction while the upper grip pulls the fabric and reduces the internal diameter of the aluminum warp. (b) The anvil applies compression locally at the mid-section of the stent. (c) The stent is globally compressed between two steel compression plates.}
	\label{Fig_2}
\end{figure}

 The experimental setup for the  \emph{in vitro} tests are shown in~\fref{Fig_2}. An environmental chamber was used to maintain the ambient temperature at $37^o$ centigrade to simulate the body temperature. The radial pressure (i.e., circumferential resistive pressure) measured by the radial crimping test according to~\citep{Morris2016,duda2000} shown in~\fref{Fig_2}(a). An aluminum fabric of width equal to the undeployed length of the stent is wrapped around the fully deployed stent and threaded through a narrow gap between two rollers (diameter of $3~mm$).The lower edge of the fabric is attached to the fixed jaw while the upper edge is attached to the moving jaw and the load cell. As the upper jaw moves upward, the  circumference of the aluminum wrap decreases leading to reduction of the internal diameter and radial crimping of the stent. Note that in this case we only measure the circumferential resistive force, not the chronic outward force that is applied by the stent to the wall during deployment. This is due to the fact that increasing the patency is commonly performed by immediate angioplasty after venous stenting. Hence, even if the chronic outward pressure is not enough, the angioplasty using balloon expansion can assist during deployment. Accordingly, the circumferential resistive force, resisting the post-deployment recoil, is a more representative characteristic in terms of  clinical durability. In a temperature control chamber, the global collapse and local collapse tests, based on the deformation modes suggested by~\citep{Dumoulin2000,Bandyopadhyay2013}, were performed by compressing the stent by rigid plates and an anvil (tip diameter of $10~mm$), respectively (See \fref{Fig_2}(a) and \emph{b}). Results from the experiments will be presented and compared with semi-analytical model in Sections 4 and 5.

\section{Unit-cell study and finite element analysis}
\begin{figure}[!ht]
	\centering
	\includegraphics[width=0.6\textwidth]{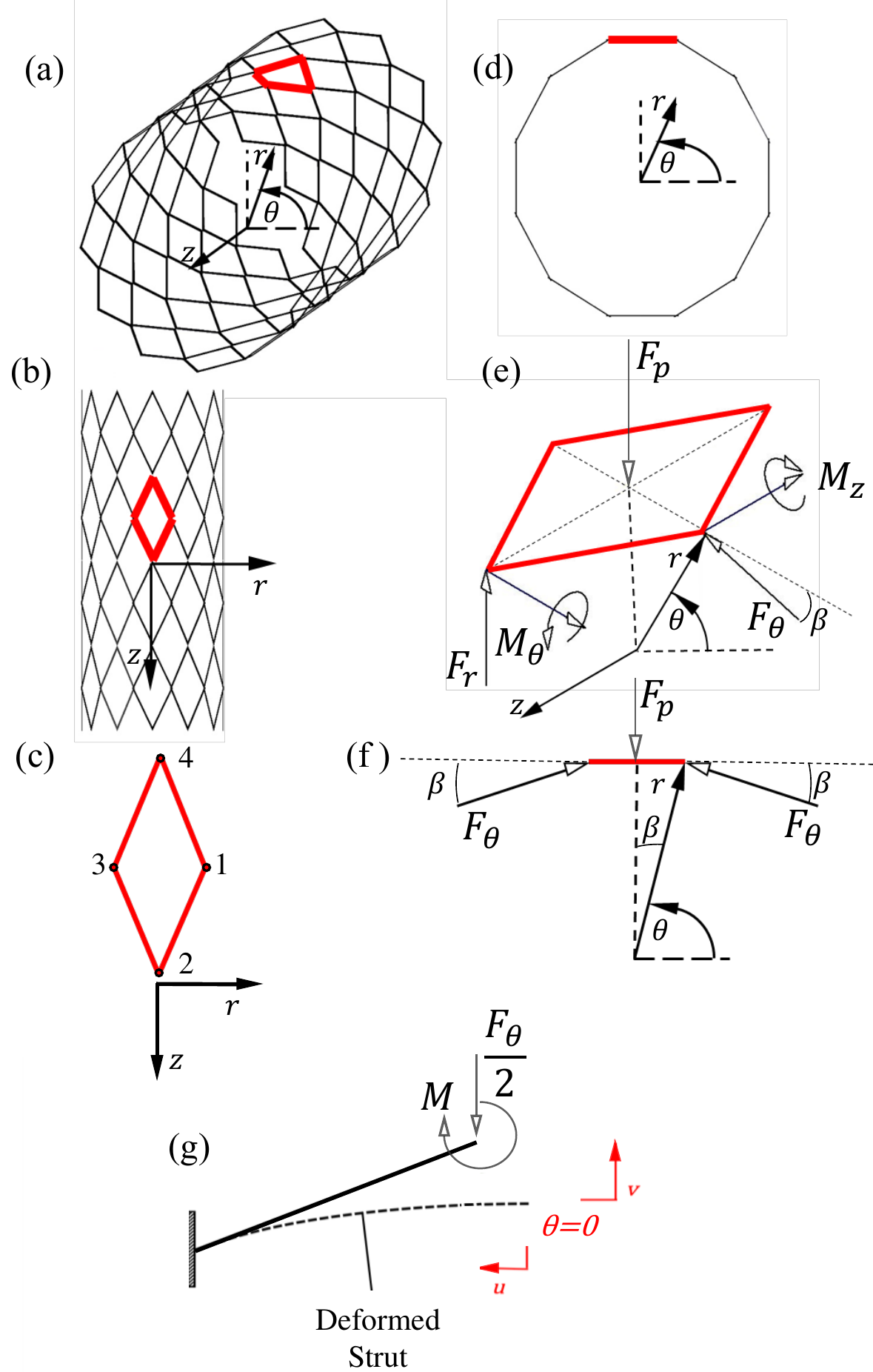}
	\caption{Typical geometry and loading conditions of a unit-cell in a stent structure. (a) cylindrical coordinate system; (b) top view of the isolated (highlighted) unit-cell in the structure; (c) four isolated joints of the unit-cell; (d)  front view of the isolated unit-cell in the structure; (e) the reaction moments and forces applied to the isolated joints, $F_p$ (resultant force due to contact pressure applied by the vessel to the stent), $M_z$ and $M_\theta$ (the reaction moment along $z$ axis and $\theta$ axis), $F_r$ and $F_\theta$ (the reaction forces along $r$ axis and $\theta$ axis); (f) the front view of a unit-cell loading condition; (g) kinematic role of a strut in deformation of a stent unit-cell. Each stent has $n_a$ number of unitcells along the axial ($z$) direction and $n_c$ number of unitcells in the circumferential ($\theta$) direction.}
	\label{Fig_3}
\end{figure}

The deformation characteristics of a stent rely on its lattice expansion mechanism, which can be investigated through a unit-cell study. This method can reduce the computation time and cost since the entire stent is not modelled. First, we start with defining the unit-cell for each design (see~\fref{Fig_3}), where a cylindrical polar co-ordinate system ($r-\theta-z$) is introduced in~\fref{Fig_3}(a). The forces and moments associated with these co-ordinate axes are $F_r$, $F_\theta$ and $M_z$, $M_\theta$, respectively. Each stent has $n_a$ number of unitcells along the axial ($z$) direction and $n_c$ number of unitcells in the circumferential ($\theta$) direction.  Assuming  periodic boundary conditions and axisymmetry of the structure, we can identify a unit-cell and define the boundary conditions at the decoupled joints/links as shown in~\fref{Fig_3}{(c)}. For uniform expansion and axisymmetric boundary conditions, the force $F_r$, and the moments $M_\theta$, and $M_z$ can be neglected~\citep{Hejazi2018}. Since Braided design is made through braiding, we cannot define a closed-joint uni-cell. Consequently, the approach to study the expansion of this stent will be different and discussed separately.

The venous pressure distribution acting on the stent is assumed to be uniform. Consequently, the resultant force $F_p$ due to pressure, which is applied by vessel on all struts, acts at the center of the unit-cell in a radial direction and it is related to $F_\theta$ (the tangential force applied to the joints) as: 
\begin{equation} \label{e1}
F_p=2F_{\theta}\sin{\beta}, \quad \beta=\frac{\pi}{n_c}.
\end{equation}
We can define the  length ($L$) and the diameter ($D$) of an expanding stent at each stage of deployment by following the method introduced by~\citep{Douglas2014} below:
\begin{equation}\label{e2}
L=n_a (l_0-2u),
\end{equation}
\begin{equation} \label{e3}
D=\frac{n_c (w_0+2v)}{\pi},
\end{equation}
where $u, v, l_0$, and $w_0$ are respectively axial displacement, circumferential displacement, undeformed length, and width of a unit-cell (\fref{Fig_3}(g)). Note that for an unexpanded stent  $u=v=0$ so that  $L=n_al_0$ and $D=\frac{n_cw_0}{\pi}$ define the initial length and diameter, respectively.

The main purpose of deploying a stent is to maintain the patency of the lumen. A vessel that tends to recoil applies a redial pressure to the stent, which governs post-deployment performance~\citep{Duerig2000,Morlacchi2013}. This pressure can be defined as:
\begin{equation} \label{e4}
P=\frac{F_p}{D\beta(l_0-2u)},
\end{equation}
Where $P$ is the lumen radial pressure (circumferential resistive pressure), the numerator is the total applied force, and the denominator is the circumferential area of a unit-cell. By substituting~\req{e1}, and~\req{e3} into~\req{e4} we have:
\begin{equation} \label{e5}
P=\frac{2F_{\theta}\sin{(\frac{\pi}{n_c})}}{(w_0+2v)(l_0-2u)}.
\end{equation}
We introduce the foreshortening parameter ($f$) as 
\begin{equation} \label{e6}
f=\frac{l_0-l}{l_0}=\frac{2u}{l_0},
\end{equation}
where $l=l_0 -2u$ is the length of the unit-cell at a given stage of deployment. Using~\req{e6}, we can rewrite the radial pressure as a function of foreshortening as follows: 
\begin{equation} \label{e7}
P=\frac{2F_\theta\tan{(\frac{\pi}{n_c})}}{(w_0 -2v) l_0 (1+f)}.
\end{equation}
The above indicates that $P$ and $f$ are inversely related. This suggests that while a large value of foreshortening is undesirable for precise deployment a larger radial pressure can be achieved. It is worth noting that this compromise in clinical performance can be avoided by choosing zero foreshortening stent-designs proposed in~\cite{Douglas2014}

Compliance of a stent is defined according to
\begin{equation} \label{e8}
C=\frac{D_2-D_1}{D_1(P_2-P_1)},
\end{equation}
where $D_i$ and $P_i$ are the incremental stent diameter and pressure. By substituting equations (2) and (3) into (8), we have
\begin{equation} \label{e9}
C=\frac{v_1-v_2}{\sin{(\frac{\pi}{n_c})}\left(\frac{(w_0+2v_1)F_{\theta_{2}}}{(w_0+2v_2)(l_0-2u_2)}-\frac{F_{\theta_{1}}}{l_0-2u_1}\right)}.
\end{equation}

Equations (2) to (9) indicate that for calculating the deformation characteristics (radial pressure, compliance, and foreshortening) we need to correlate the strut bending force ($F_\theta$) with displacements in the circumferential ($v$) and longitudinal ($u$) directions. In Sections 3.1 and 3.2 we study the bending mechanism of the candidate stents, which governs the deformation characteristics of the stent.

\subsection{Unit-cell deformation characteristics of Chevron design and Diamond design }
\begin{figure}[!ht]
	\centering
	\includegraphics[width=\textwidth]{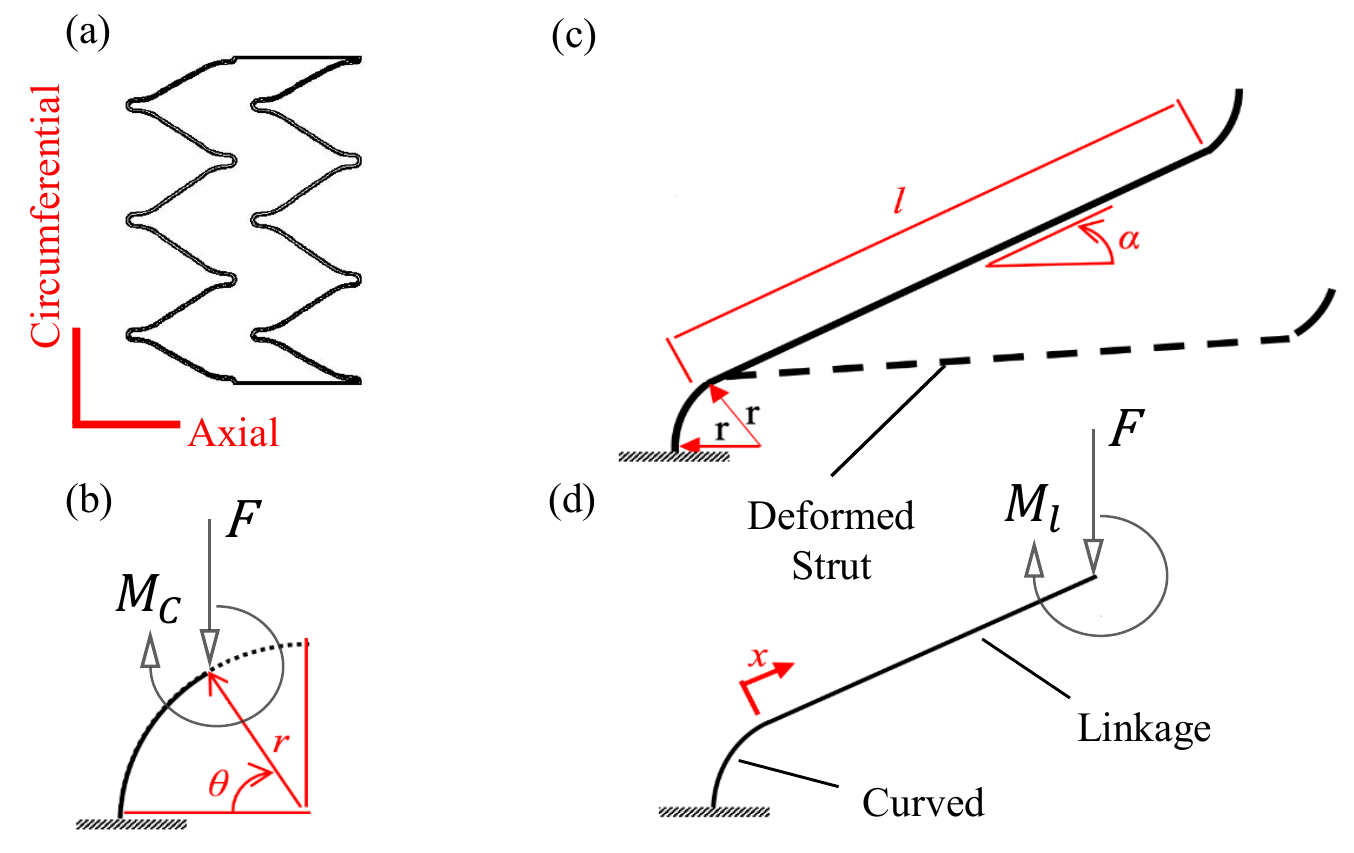}
	\caption{(a) Chevron design unit-cell geometry; (b) a strut of the unit-cell including the curved and linear parts; (c) the bending moment and shear force at a given cross section of the curve part; (d) the bending moment and shear force at a given cross section of the linkage.}
	\label{Fig_5}
\end{figure}

Chevron design and Diamond stent designs used in this study are made of Nitinol alloy, which provides the desired mechanical properties such as super-elasticity. The Nitinol struts undergo phase transition depending on the mechanical strain which influences their deformation characteristics. A mathematical model is introduced for the bending analysis of Nitinol beams in~\citep{Mirzaeifar2013} .  We apply their model to find the deformation of the stent strut subjected to bending described in~\fref{Fig_3}(g). The bending analysis for a cantilever beam is summarized in Appendix A1. The Chevron design  has a more complex shape, which can be divided into curved and straight sections (\fref{Fig_5}). Hence, we have to modify~\req{e5},~\req{e6},~\req{e7}, and~\req{e9} by substituting circumferential displacement $v$ by $3v$ and longitudinal displacement $u$ by $0.5u$ to account for the number of struts in the unit-cell and the different orientation of the joints.  We can use equations (10) and (11) respectively, to calculate the bending moment in the curved and straight portion as:
\begin{equation} \label{e10}
M_c=\frac{1}{2}Fl\cos{\alpha}-Fr(1-\cos{\theta}),
\end{equation}
\begin{equation} \label{e11}
M_l=\frac{1}{2}F l\cos{\alpha}-Fx\cos{\alpha}+Fr.
\end{equation}
where $\theta$ and $x$ locate the section in the curved and the straight part, respectively in ~\fref{Fig_5}(b) and in~\fref{Fig_5}(d). Using ~\req{e10} and~\req{e11} we can calculate the bending moment throughout the strut of Chevron design as a function of $\theta$ for the curved portion and $x$ for the straight part. In these equations, $l$ is the effective arm length of the strut (contributing in bending moment), $r$ is the radius of the curved portion, and $F=F_{\theta}\cos\beta$ based on~\fref{Fig_5}.  For the Diamond stent (Nitinol) since there is no curved region at the intersection of joints, we can directly use the analysis of a cantilever beam in Appendix A1 to calculate the internal bending moments during deployment. Having thus found the internal forces we can calculate the deformation properties of these two designs by following the Appendix A1 to find the displacements and calculate radial pressure and compliance using~\req{e7} and~\req{e9} .

\subsection{Unit-cell deformation characteristics of Z design and Braided design}
 Geometric design imparts self-expanding ability to Z design and Braided design stents. Two sets of steel links and a coil that connect these links are the elements that define the Z design unit-cell. Accordingly, we can calculate the displacement of the strut based on the coil angular twist and the elastic bending of the link. The displacement in a single strut is a combination of the coil angular twist and the link bending deflection. The torsional stiffness ($k_t$) of the coil and the torsion angle can be calculated through $k_t=\frac{Ed^4}{10.8Dn}$, where $E, d, D, n,$ and $l$ are torsional stiffness, Young's modulus, wire diameter, the diameter of the coil, number of coil body turns, the torsion angle and link length~\citep{Budynas2008}. Accordingly, the angular twist of the coil part, longitudinal and circumferential displacements can be determined through    
\begin{equation}\label{e12}
\gamma=k_{t}F(l+2d),
\end{equation}
\begin{equation}\label{e13}
u=l(1-\cos{\gamma}),
\end{equation}
\begin{equation}\label{e14}
v=\frac{Fl^3}{3EI}+l\sin{\gamma}.
\end{equation}

In~\req{e12} we can determine the longitudinal displacement of the strut tip in terms of the angular twist. Here, we ignore the contribution of the strut bending as it is in the elastic region of the bending deformation and can be assumed small in comparison to the effect of an angular twist. To calculate the circumferential displacement, we can use~\req{e13}, in which, the first term is the contribution of bending displacement and the second term represents the effect of coil angular twist. 

Braided design is fabricated by braided wires forming its entire structure. Consequently, it does not have any true geometrical unit-cell or joint and therefore no real strut can be defined. The expansion of Braided design was investigated to derive an equation that relates the pressure to the diameter of the Braided design structure based on slender bar theory~\citep{Wang2004a,Wang2004b} as:
\begin{equation} \label{e15}
P=\frac{n\cos^2{\alpha}}{2\pi{r^2}\sin^2{\alpha}}\left[ \begin{split} &\frac{EI\sin{\alpha}}{r}\left(\frac{\cos^2{\alpha}}{r}-\frac{\cos^2{\alpha_0}}{r_0}\right)\\ &-\frac{GI_{p}\cos{\alpha}}{r}\left(\frac{\cos{\alpha}\sin{\alpha}}{r}-\frac{\cos{\alpha_0}\sin{\alpha_0}}{r_0}\right)\end{split} \right],
\end{equation}
\begin{equation}\label{e16}
f=\frac{\sqrt{\lambda_{0}^2+4\pi^2r_{0}^2-4\pi^2r^2}-\lambda_{0}}{\lambda_{0}},
\end{equation}
Where $n, E, G, I_p, r,$ and $r_0$ are respectively number of wires, Young's modulus, shear modulus, the area moment of inertia, stent radius, and nominal radius. To calculate the foreshortening, we can use~\req{e16}, in which $\lambda_0$ and $r_0$ are initial helical wire pitch and diameter. Furthermore, to determine the compliance, instead of using~\req{e9}, which has been used for other stents, we can directly use~\req{e8} associated with radial pressure values from~\req{e15}. 

\subsection{Finite element simulation parameters and material properties}

In this work, the simulation has been performed in ABAQUS/Standard commercial code linked with a user material subroutine (UMAT) based on~\citep{Lagoudas2008}. Here, the FE method was employed to evaluate two different problems. We studied the bending of the Chevron design and Diamond design struts to validate the  method presented in~\citep{Mirzaeifar2013}, which addresses the bending mechanics of a Nitinol beam. We used material properties of NITI-I for Nitinol stents, and an elastic modulus of 193 GPa, Poisson's ratio of 0.3 and uni-axial yield stress of 260 MPa for steel stents~\citep{Bandyopadhyay2013,Duerig2000}. We employed UMAT (user material subroutine), which is a framework for ABAQUS users to implement a material (Nitinol model was not available in the software library at the time of this study). The foundation of the UMAT code is the thermo-mechanical constitutive model of Nitinol~\citep{Auricchio1997,Bhattacharya2003,Lagoudas2008}. We used fully integrated solid linear hexahedron element (C3D20) for other designs. The global size for the meshing varies from 0.075 mm to 0.05 mm based on the mesh sensitivity test. The numerical finite element  results are used to validate the semi-analytical method for bending of Nitinol struts, presented in Section 3.1.

\section{Results}

\begin{figure}[!h]
	\centering
	\includegraphics[width=\textwidth]{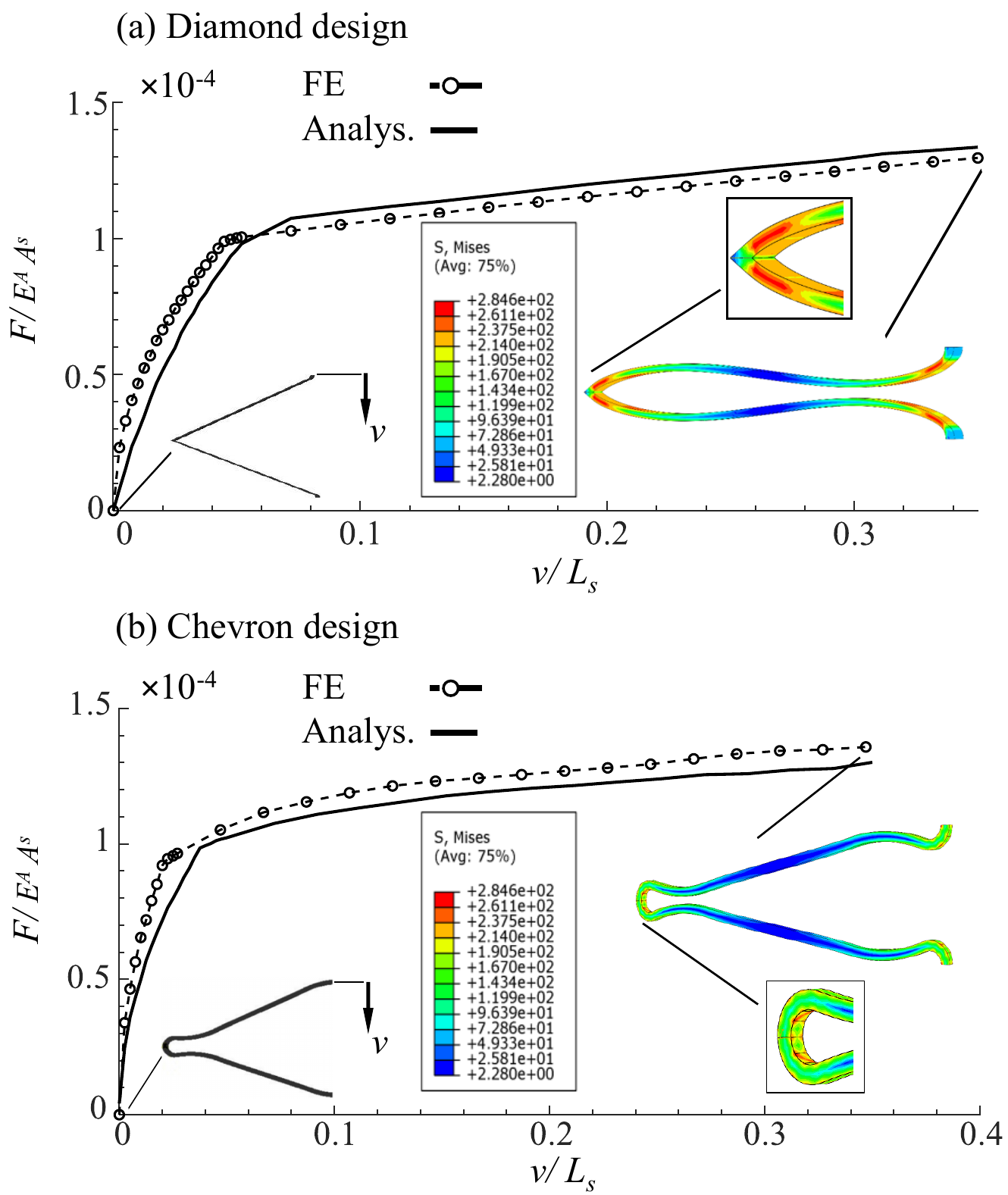}
	\caption{Von Mises stress distribution over the stent strut at the delivery size (maximum magnitude of stress); (a) Diamond design joint; (b) Chevron design. Comparison between strut bending force vs. circumferential displacement from the analytical method in Section 3 and Finite Element calculations.}
	\label{joint}
\end{figure}

A comparison is drawn between FE simulation (Section 3.2) and the semi-analytical approach (Section 3.1) results for a strut bending of the Nitinol stents in~\fref{joint}. The change of the slope reflects the onset of phase transition (austenite to martensite). The FE solution here is stiffer for Chevron design. For Diamond design, however, the semi-analytical method demonstrated a stiffer response. This is because of the shear stress contribution to the phase transition, which has been ignored in Section 3.1. The change of the slope reflects the onset of phase transition (austenite to martensite). If von Mises stress is more than 260 MPa, the region has a pure martensite phase. A core of pure austenite always exists for the Chevron design joints. For Diamond design, however, we observe a core of martensite phase close to the joints.

\begin{figure}[!ht]
	\centering
	\includegraphics[width=\textwidth]{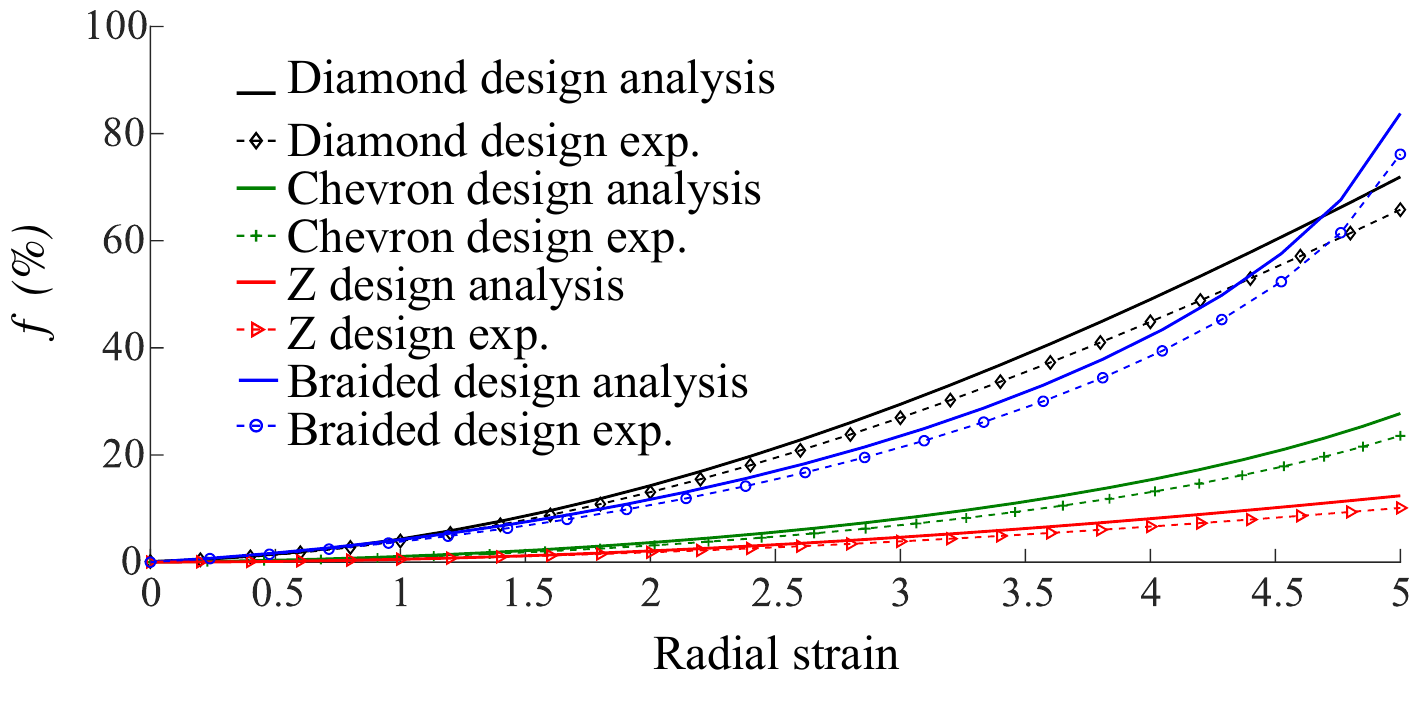}
	\caption{Analytical prediction of foreshortening compared with experimental measurements.}
	\label{FS}
\end{figure}

The foreshortening of the stents, illustrated in~\fref{FS}, has been calculated through~\req{e6}) for joint based designs (Z design, Diamond design, and Chevron design) and~\req{e16} for Braided design. It should be noted that the foreshortening is a dimensionless characteristic and the presented results in~\fref{FS} are valid for different stent sizes. As shown, in all cases the experimental values of foreshortening is smaller. This can be the result of longitudinal compressive forces (due to the friction) that is applied to the stent during the crimping test in the aluminum fabric.   

\begin{figure}[!ht]
    \centering
    \includegraphics[width=\textwidth]{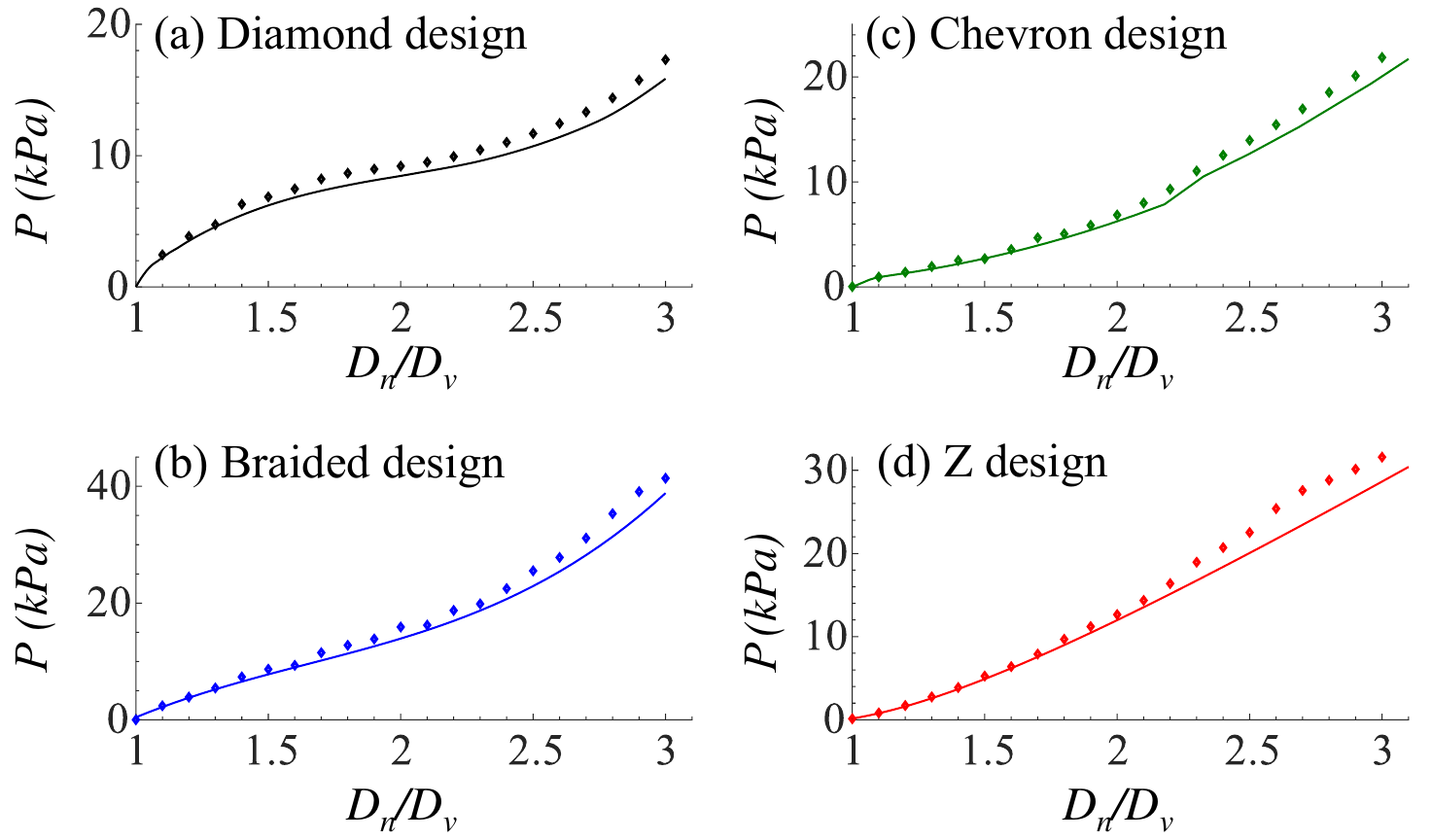}
    \caption{Radial pressure calculated based on~\req{e5}, shown by solid line, compared with data points corresponding to \emph{in vitro} experiment (Fig \fref{Fig_2}(a)). $D_n$ and $D_v$ are stent nominal expanded diameter and vessel diameter, respectively. Usually the ratio $\frac{D_n}{D_v}$ does not exceed 1.3 which corresponds to 30\% oversizing.}
    \label{PFA}
\end{figure}

In  clinical practice, stents are oversized to maintain a required radial pressure.~\fref{PFA} shows the radial pressure versus the over-sizing parameter ($D_{n}/D_{v}$), where $D_n$ and $D_v$ are stent nominal expanded diameter and vessel diameter, respectively. For a given $D_{n}=16mm$, the experimental data points were measured through the  \emph{in vitro} test setup shown in Section 2. In this case, we simulate $D_{v}$ by adjusting the diameter of the aluminum wrap. For each stent, the joint analysis (Sections 3.2 and 3.3) yields the circumferential force ($F_\theta$) and longitudinal displacement ($u$), at a given circumferential displacement ($v$). In this case, circumferential displacement is adjusted to match the simulated vessel diameter using~\req{e3}. Accordingly,~\req{e5} was used to calculate the radial pressure.

\begin{figure}[!ht]
	\centering
	\includegraphics[width=0.85\textwidth]{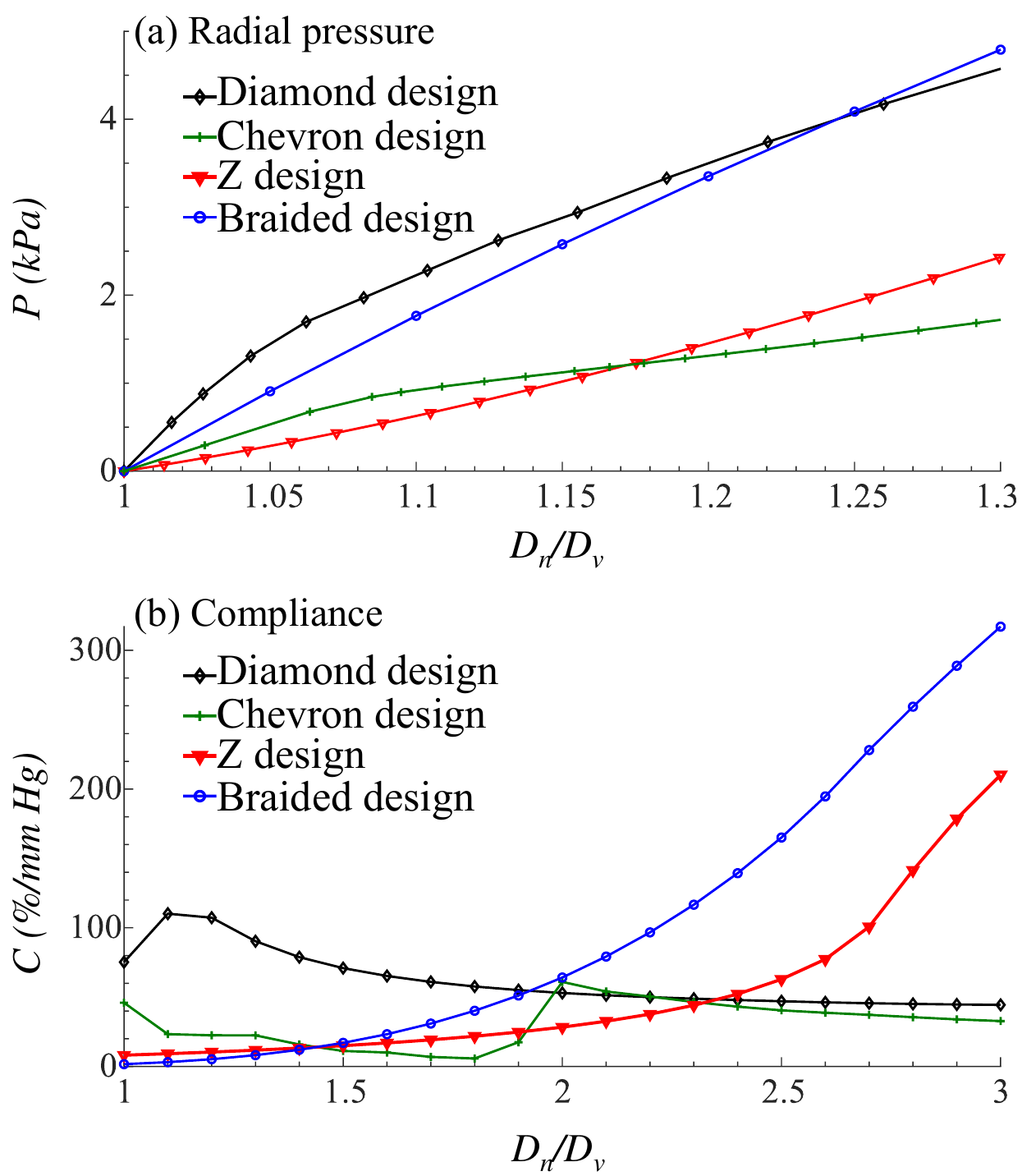}
	\caption{Experimental measurements of radial pressure and compliance of stents compared in a single plot.}
	\label{Comp}
\end{figure}

We can compare the radial pressure performance of the stents in~\fref{Comp}(a), where we combine and compare all stents in~\fref{PFA} in a single plot for a better comparison. Note that the oversizing parameter is limited to $D_n / D_v=1.3$ since 30\% oversizing is recommended in most of the clinical cases. Diamond design and Braided design have a larger radial pressure in this range. Although, Z design has the highest maximum radial pressure at $D_{n}/D_{v}=3$, it loses 70\% of its radial pressure at $D_{n}/D_{v}=1.5$. To compare stent radial compliance, calculated using~\req{e8}, we have~\fref{Comp}(b). The steel stents (Braided design and Z design) are much more compliant at higher oversizing values. However, in the range of $D_{n}/D_{v}<1.5$, they are much stiffer. The compliance of Nitinol stents, Chevron design and Diamond design, does not change as much as steel stents with changing the oversizing $D_{n}/D_{v}$. This can be due to the effect of the phase transition, which is evident in the local maximum points in the compliance curve. 

\begin{figure}[!ht]
    \centering
    \includegraphics[width=0.9\textwidth]{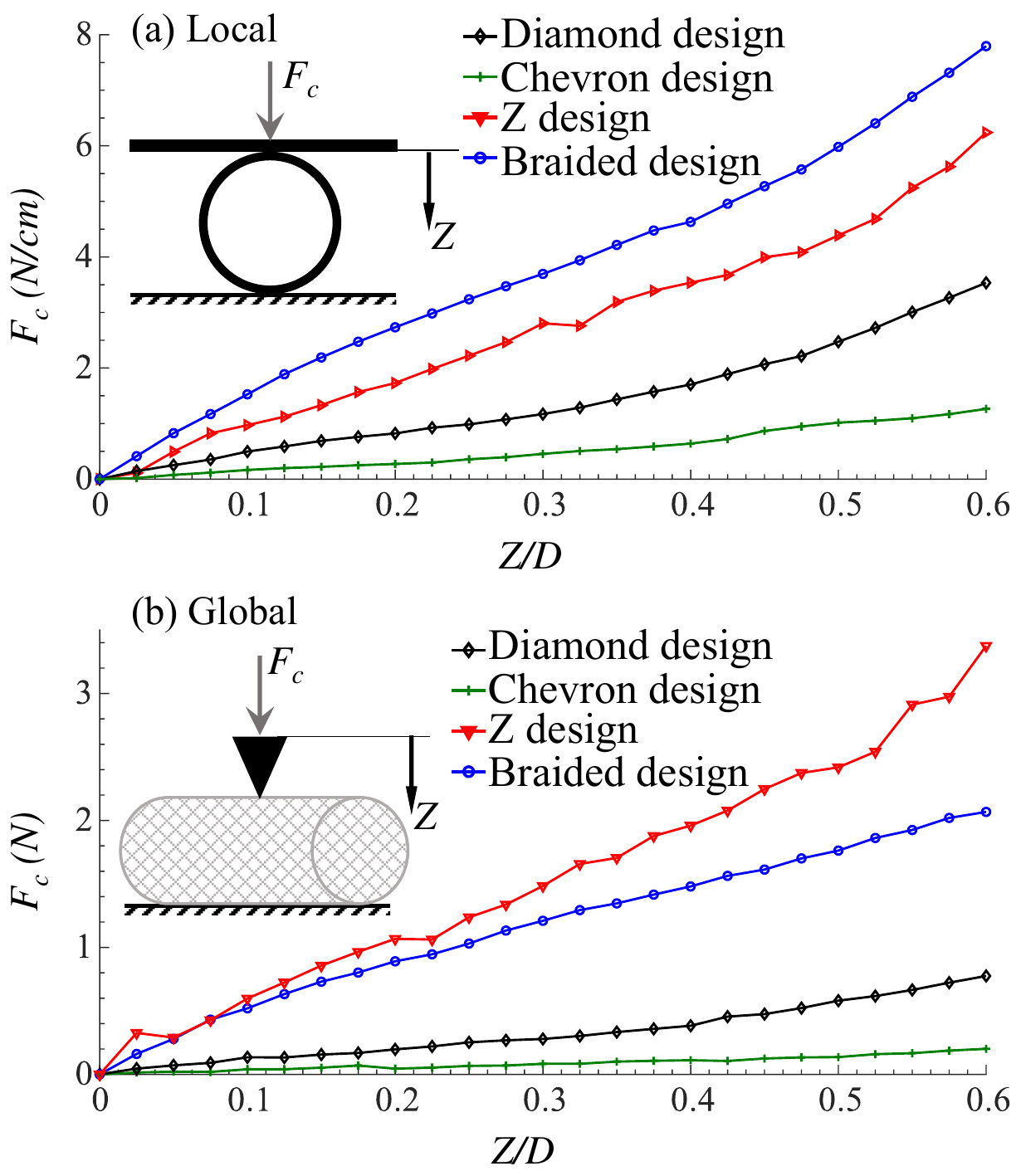}
    \caption{Experimental measurements of collapse force ($F_c$) Vs. displacement ratio ($Z/D$). $D$ is the internal diameter of the fully expanded stent.}
    \label{Col}
\end{figure}

The results of the global and local collapse tests are shown in~\fref{Col}(a) and (b), respectively. Wall stent and Z design have the higher resistance in both tests compared to Nitinol stents. Note that Z design has a higher resistance to localized collapse while Braided design performs better in global collapse test. It is worth mentioning that steel stents were also stiffer in the radial pressure test when $D_{n}/D_{v}<1.5$.   

\section{Discussion}
\subsection{Results assessment, validation, and limitations}

The assumption on loading conditions and resultant deformation of the semi-analytical method is validated through comparison with experiments (\fref{PFA}). The experiments show a slightly higher pressure, especially at higher expansion ratio ($\frac{D_n}{D_v}>1.3$).
This is partly due to friction forces exerted by the wrapping foil used in experiments, and the loading-unloading hysteresis.  Further, in the semi-analytical method pure bending is assumed and torsional loads are ignored as axi-symmetric deformation is assumed.  The role of shear in the deflection of the strut accounts for a portion of error, particularly for Diamond design as we can see in~\fref{joint}, where FE predicts the phase transition at a smaller displacement. The curved part of the Chevron design structure is connected to the straight strut by another fillet curve. Since the connection between these parts is assumed as a straight link for the purpose of simplification, the results deviate from the experiments. This conclusion is also valid for the fillet that connects the curved and straight parts of the Z design (See~\fref{Fig_5}).

In ~\fref{Comp}(a), we observe a higher radial pressure (up to 30\% oversizing,$D_n / D_v=1.3$) for Diamond design and Braided design (steel stents).  They offer  more scaffolding than Z design and Chevron design (Nitinol stents) due to higher coverage. The radial pressure is a function of bending force in each unit-cell, and the bending force itself is a function of the cross-section area, length of the strut, the material of the stent, and foreshortening according to~\req{e7}. Hence, with higher foreshortening, the material volume per unit length increases and the total contact surface decreases, which leads to higher radial pressure. Another remarkable observation is the relationship between compliance and collapse resistance.  By comparing~\fref{Comp}(b) and \fref{Col}, we can conclude that steel stents are less susceptible to collapse. Consequently,  collapse resistance is inversely correlated to compliance. When the collapse mode has a spatially non-uniform deformation as a result of structural instability  the analytical method given here is not applicable. However, given the relationship between collapse resistance and radial compliance (calculated based on~\req{e9}), it is possible to qualitatively compare the collapse behavior of stent designs.  

The limitation of the present study arises in both experimental and analytical modelling. The  \emph{in vitro} experiment based on Section 2 only considers the uniform deployment and does not account for non-uniform lumen or a curved anatomy of the vein. In general it is a challenge to excise fibrotic veins  and the vein material properties are difficult to emulate using polymeric tubes. This means that the vein-stent interaction is not accounted in this study. This is an area where further work is needed. However, for a given vein model the relative performance that we report in this study is expected to hold. Another alternative approach to measure the radial pressure is to use an aperture-type (crimper) machine (see~\citep{dabir2018,mckenna2020}. The limitations listed above are still unavoidable. Another consideration is the effect of foreshortening on the radial pressure test. In the setup used here, it is easy to use a wider fabric to account for stent elongation during the crimp test. In the aperture-type devices, however, the length of the stent is limited to the device capacity. In the analytical approach (Section 3), we ignored the frictional forces and the mechanical interaction between the vein wall and the stent was modelled as a uniform pressure distribution, which is not the case for curved vessel geometries. This effect can be included in future studies by assuming the elasticity of the vessel wall.  

\subsection{Clinical significance}
This study compares four different venous stent designs (Chevron design, Z design, Diamond design, and Braided design) based on their collapse, foreshortening, radial pressure, and compliance. Usually, the stent deployment for abnormal/diseased vein is performed after the vein angioplasty under fluoroscopy. Consequently, a surgeon can observe the length of occlusion, the regions with high forces (based on the shape of inflated balloon), and the stiffness of the occluded part by observing the balloon pressure. Based on the mechanical properties of the diseased vein, we can suggest the most suitable design for a specific occlusion type. Based on our study, here we present a summary of some of the common vein occlusion scenarios.

\begin{itemize}
    \item Z design has superior performance against collapse deformation. Consequently, it may be more reliable to treat diseases like May-Thurner syndrome, which tends to apply localized force.
    \item Nitinol stents are more compliant. Thus they follow the joint movements. Accordingly, they are more suitable for deployment in proximity of expected major vessel bending during limbs movements. Diamond design and Braided design apply a higher radial pressure. Thus, they may be chosen for the long lesion with high recoil. 
    \item Stents have a number of anchors at both ends, which attach them to the vessel wall to avoid migration. If the locations of the stent tips are critical to be predicted (e.g., deployment close to branch orifice), Z design can be a good choice. Because of the small foreshortening of Z design in comparison to other designs, we can predict the final location of the ends. 
\end{itemize}
\section{Conclusions}
This study compared four different stents currently used in veins. Two designs (Z, braided stents) are off-label while the remaining two (Diamond and Chevron) are specifically designed for venous stenting.  Deformation characteristics of all four designs are compared under identical loading conditions through \emph{in -vtro} testing and semi-analytical modelling. Particular attention is given to foreshortening, compliance,  radial pressure, and collapse resistance. An inverse correlation between radial compliance and collapse resistance, and foreshortening and radial pressure is found. A good agreement  is found between the predictions  of the unit cell based semi-analytical modelling and experiments. Relative merits of each stent design for common vein occlusion scenarios are identified. Venous stenting is a relatively new area of investigation compared with arterial stents. While we expect the relative comparision across the designs to hold, further work on vein-stent interaction is needed.

\section*{Conflict of interest}
The authors claim no conflict of interest regarding the choice of candidate stents. No human or animal test was conducted for this study.
\section*{Acknowledgment}
We sincerely acknowledge the Division of Vascular Surgery at Vancouver General Hospital for providing the stent models. Furthermore, the funding through a Discovery Grant to Srikantha A. Phani from Natural Sciences and Engineering Research Council (NSERC) Canada is acknowledged.  

\bibliography{jbmbib}

\section*{A1. Material properties and Elastic bending of Nitinol beams}
\begin{figure}[!ht]
	\centering
	\includegraphics[width=0.9\textwidth]{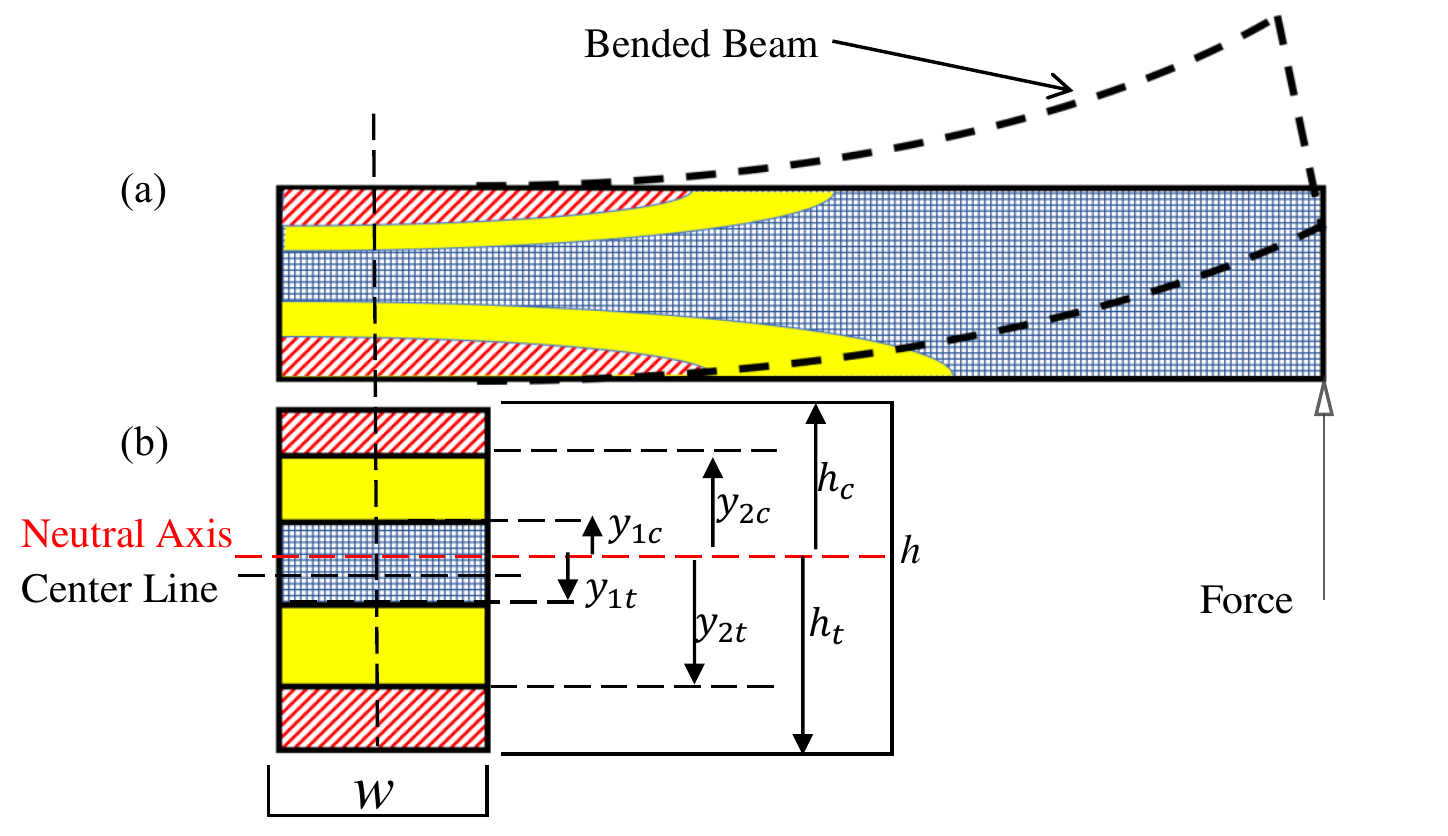}
	\caption{Phase distribution of a Nitinol beam under bending.; (a) phase distribution, blue (meshed), yellow (solid) and red (dashed) are respectively, Austenite, Transition and Martensite Phases; (b) Cross section stress distribution.}
	\label{Fig_A1}
\end{figure}

The correlation between elastic curvature ($\kappa$) and the bending moment is given in~\req{e17}, where $y_{1c}$ and $y_{2c}$ are respectively the distance from the neutral axis to the boundaries of the transition, and martensite regions, which are under compression (\fref{Fig_A1}); the same terminology is true for $y_{1t}$ and $y_{2t}$ which are representing the same variables for the tension portion. Thermo-mechanical properties for NITI-I are given in Table \ref{T_A1}. To calculate $I$ in~\req{e17},~\citep{Mirzaeifar2013} introduced four different functions for different loading conditions. Here, we used equation (24) in their paper. Accordingly, we determine the bending moment in each strut (a function of $F_\theta$) and find the curvature of the stent strut based on the method presented by~\citep{Mirzaeifar2013}. Once we have the curvature along the strut length, we can find the circumferential and longitudinal displacements.
\begin{multline} \label{e17}
M(\kappa)=-1/3  E^A \kappa(y_{1c}^3-y_{1t}^3 )+(I(y_{2c})-I(y_{1c}))+\\ E^M w[1/3 \kappa((h_{c}^3)/8-y_{2c}^3)-H^c((h_{c}^2)/4-y_{2c}^2)]+(I(y_{2t})-I(y_{1t}))\\ +E^Mw[1/3\kappa(y_{2t}^3-(h_{t}^3)/8)-H^c(y_{2t}^2-(h_{t}^2)/],
\end{multline}
Where $y_{1c}, y_{2c}, y_{1t},$ and $y_{2t}$ are given in equations (32) in~\citep{Mirzaeifar2013}. Other parameters in the above expression are given in the Table~\ref{T_A1}.
\begin{table}[!ht]
	\centering
	\begin{tabular}{ll}
		\hline
		Stainless Steel Stents&\\
		Elastic modulus&193 GPa  \\
		Poisson ratio&0.3\\
		Uni-axial yield stress&260 MPa\\
		&\\
		Nitinol Stents~\citep{Bandyopadhyay2013}&\\
		$E^A$&72 GPa\\
		$E^M$&30 GPa\\
		Poisson ratio&0.42\\
		$H^c$&-0.035 \\
		\hline
	\end{tabular}
	\caption{Material properties for the stents used in this study.}
	\label{T_A1}
\end{table}

\end{document}